\documentclass[fleqn,twoside]{article}
\usepackage{espcrc2}   

\usepackage[cp850]{inputenc}
\usepackage{verbatim}

\usepackage{graphicx}
\usepackage[figuresright]{rotating}

\newcommand{\AmS}{{\protect\the\textfont2
  A\kern-.1667em\lower.5ex\hbox{M}\kern-.125emS}}

\title{Finite Size Scaling, Fisher Zeroes and $\cal{N}$=4 Super Yang-Mills
\vspace{0.2cm}}

\author{P.R. Crompton\address
  {Institut\ f\"ur Theoretische Physik, Universit\"at Leipzig, Augustusplatz 10/11, D-04109 Leipzig, Germany.
    \vspace{-0.6cm}},
  W. Janke\addressmark,
  Z.X. Xu\address{Zhejiang Institute of Modern Physics, Zhejiang University, Hangzhou 310027, People's Republic 
    of China.} and 
  H.P. Ying\addressmark \vspace{0.2cm}}

\begin{document}

\begin{abstract}
We investigate critical slowing down in the local updating continuous-time Quantum Monte Carlo 
method by relating the finite size scaling of Fisher Zeroes to the dynamically generated gap, 
through the scaling of their respective critical exponents. As we comment, the nonlinear 
sigma model representation derived through the hamiltonian of our lattice spin model can also be used to give a 
effective treatment of planar anomalous dimensions in $\cal{N}$=4 SYM. We present scaling arguments from our FSS 
analysis to discuss quantum corrections and recent 2-loop results, and further comment on the prospects of extending this approach for 
calculating higher twist parton distributions.  
\vspace{-1mm}
\end{abstract}

\maketitle

\section{VBS PHASE TRANSITIONS}
We investigate the critical behavior of Valence-Bond-Solid (VBS) states in quantum spin chains by means of Quantum Monte 
Carlo (QMC) simulation using the continuous-time loop cluster algorithm \cite{qmc}. Following 
\emph{Haldane's conjecture} a variety of exotic magnetic phenomena can be attributed to the formation of ground states 
with an energy-gap in integer spin systems at low temperatures. These gapped states are predicted to have transitions 
with a massless excitation between phases, and can be effectively expressed in terms of a VBS state (spin-singlet) 
picture. The \emph{Lieb-Schultz-Mattis argument} to this conjecture has led to the recent introduction of a string 
exact order parameter to characterise these VBS transitions. We investigate the Finite Size Scaling (FSS) properties 
of a generalised \emph{twist} order parameter for a periodic mixed-spin chain \cite{ying} with a unit cell of the form 
($1, 1, \frac{1}{2}, \frac{1}{2}$). Also determining the FSS properties of the complex-temperature (Fisher) zeroes of 
the partition function \cite{zeros}, evaluated in a new scheme through knowledge of the QMC transfer matrix. 
This enables us to 
separate pseudocritical and critical point scaling behavior relating to the correlation length exponent and also to 
locate the VBS state transition points through an independent and expeditious means. Leading corrections to the FSS of 
the zeroes are also known exactly for comparable models such as the 2D Ising model with Brascamp-Kunz boundary value 
conditions \cite{wzeros}. We compare the FSS of these indicators to evaluate the critical exponents for the VBS 
transitions, comparing with nonlinear $\sigma$-model predictions.
               
The Hamiltonian for the AHFM two-spin chain with spins ${\bf S}^{\bf{a}}
$ and ${\bf S}^{\bf{b}}$ of period-4 is given as, 

\begin{figure}
  \begin{center}
    \includegraphics[width=4cm]{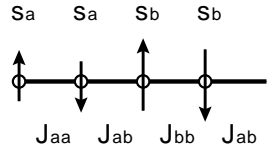}
  \end{center}
  \vspace{-13mm}
  \caption{Periodic mixed spin cell, \cite{nonlinsig}.}
\end{figure}

\begin{eqnarray}
  H && \!\!\!\!\!\!\! = \sum_{j=1}^{N/4} \, 
  ( J_{\bf{aa}} \, {\bf S}^{\bf{a}}_{4j+1} \cdot {\bf S}^{\bf{a}}_{4j+2} 
  + J_{\bf{ab}} \, {\bf S}^{\bf{a}}_{4j+2} \cdot {\bf S}^{\bf{b}}_{4j+3} 
\nonumber \\
  && \!\!\!\!\! + J_{\bf{bb}} \, {\bf S}^{\bf{b}}_{4j+3} \cdot {\bf S}^{\bf{b}}_{4j+4} 
  + \,\, J_{\bf{ab}} \, {\bf S}^{\bf{b}}_{4j+4} \cdot {\bf S}^{\bf{a}}_{4j+5} )
\end{eqnarray}

with  $J_{\bf{aa}}$=$J_{\bf{bb}}$=1 and the coupling anisotropy $\alpha = 
J_{\bf{ab}} / J_{\bf{aa}}$.

We anticipate competing low-temperature \emph{VBS} dimer gap states from 
nonlinear sigma model treatment \cite{nonlinsig}, separated via a quantum phase transition 
at critical anisotropy at $\alpha_{c}$. 
We use a Lieb-Schultz-Mattis extension \emph{twist-operator} on the groundstate $(\, |\Psi\rangle = U|\Psi_0\rangle\,\! )$  
\cite{twist}, as an exact order parameter for the VBS states to signal this transition,

\begin{figure}
  \begin{center}
    \includegraphics[width=6.5cm]{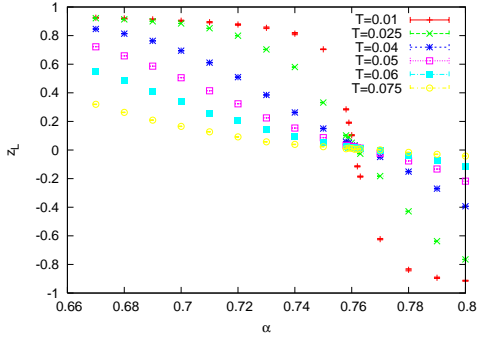}
  \end{center}
\vspace{-13mm}
\caption{Twist order parameter, $z_L$.}
\end{figure}

\begin{equation}
  {\cal{O}}_{\rm string} \!\!= \!\! -\lim_{|k-l|_{\rightarrow\infty}}
  \! \langle\Psi_0| S_{k}
  \exp\!\!\left[{\rm i} \pi\!\!\!\sum_{j=k+1}^{l-1} \!\!\!\! S_{j}\right]\!\!S_{l}|\Psi_0\rangle ,
  \label{eqn:string}
\end{equation}

\begin{eqnarray}
  U\equiv\exp\!\left[{\rm i} \frac{2\pi}{L}\sum_{j=1}^L j S_j^z\right]\!, 
  \,\,\,\,\,\,\,\,\, z_L
  \equiv\langle\Psi_0|\Psi\rangle ~,
\label{eqn:def_z}
\end{eqnarray}

Preliminary indications are of a low-temperature 2nd-order transition 
line at a (constant) $\alpha_{c}$ with a critical endpoint at $\beta_{c}=0.09$
followed by a crossover, Fig.2 and Fig.3.
              
\begin{figure}
  \begin{center}
    \includegraphics[width=6.5cm]{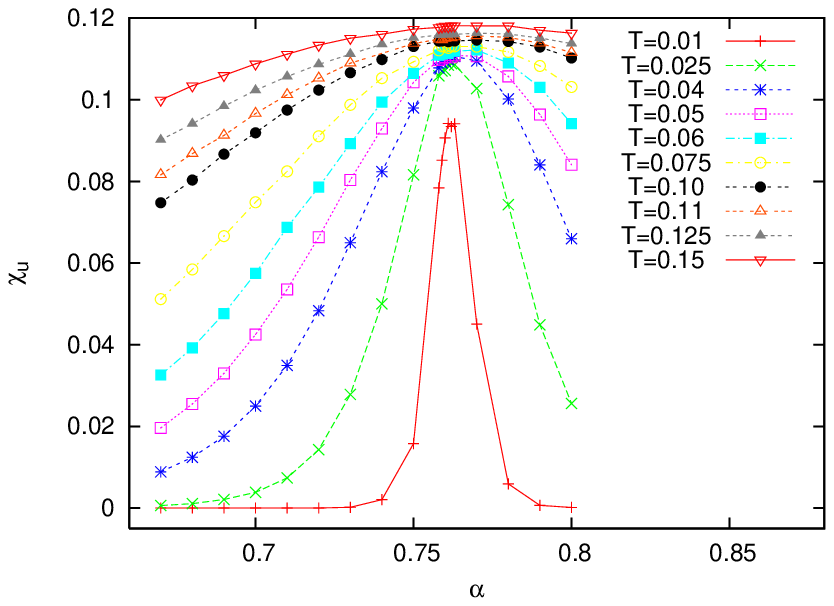}
    \includegraphics[width=6.5cm]{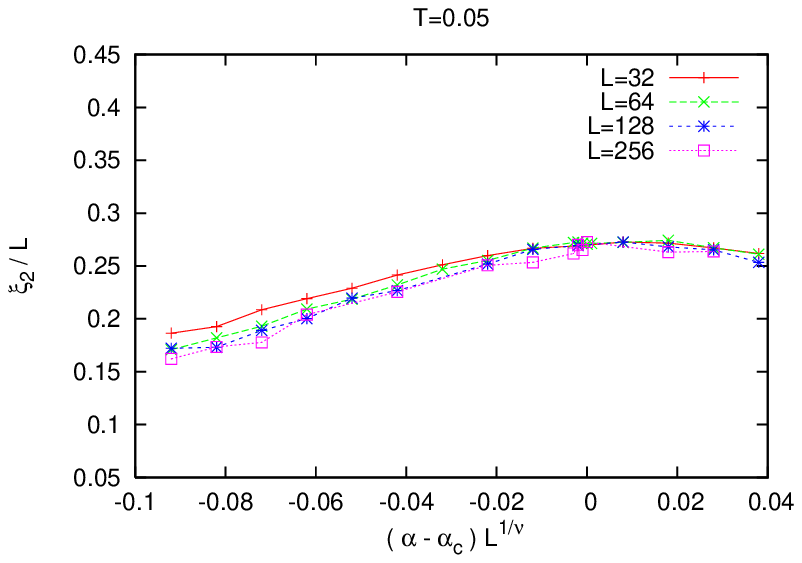}
  \end{center}
\vspace{-13mm}
\caption{Uniform susceptibility $\chi_{_{\rm{u}}}$ , and $\xi_{2}/L$.}
\end{figure}
          
\section{QMC TRANSFER MATRIX}

Following the Quantum Monte Carlo \emph{improved-estimator} definitions for 
conventional thermodynamic (2nd derivative of free energy) observables \cite{stsus} 
such as the uniform  $\chi_{_{\rm{u}}}$ and the second moment definition of correlation length $\xi_{2}$,  

\begin{equation}
  C({\bf{r}})=\frac{1}{\beta^{2}}
  \int_0^{\beta} d\tau d\tau' \big\langle\hat{S}_{{\bf{i}}}(\tau)
  \hat{S}_{{\bf{i}}+{\bf {r}}}(\tau') \big\rangle~ ,
  \label{e.defs.cr}
\end{equation}
\begin{equation}
  \chi({\bf {q}}) = \beta
  \sum_{\bf {r}}~e^{i {\bf {q}}\cdot {\bf {r}}}
  ~ C({\bf {r}}):\,\,\,\,\,\,\,\,
  \chi_{_{\rm{u}}}=\chi({\bf {q}}{=}0)~,
  \label{e.defs.chis}
\end{equation}
\begin{equation}
  \xi_{2} = \frac{L}{2 \pi}
  \sqrt{\frac{\chi (\pi,\pi)}{\chi (\pi+2\pi/L,\pi)} - 1 } ~,
  \label{secondmom}
\end{equation}
we introduce a Fisher zeros \emph{improved-estimator} polynomial expansion in the coupling $J_{ab}$ to determine 
independently the FSS properties of $\alpha_{c}$. We plot the Fisher zeros in the complex $J_{ab}^{ \, 2}$ plane in 
Fig.4, 
noting the pinching of the real axis corresponding to 
the discontinuities in the partition function that in the thermodynamic limit give $\alpha_{c}$.
The \emph{inner} edge-singularity gives the quantum phase transition (crossover) point $\alpha_{c}$, and also a clear 
indication of finite size effects where the zeroes solution 
reflected from the QMC Euclidean-time boundary conditions on our 
smallest volume ($L=64$) are lost.
The \emph{outer} edge-singularity gives us a new means to study the criticality of the Spontaneous Symmetry Breaking 
associated with the 
dimensional reduction of the QMC approach (akin to the $1D$ Ising model, where a nonzero 
magnetisation gives a non-analyticity in zeroes on the unit circle). 
Note the volume-dependent deviation from unity for imaginary valued 
($J_{ab}^{ \, 2} < 0$) critical couplings with relation to the \emph{twist-operator}. Although evidently there is no VBS quantum 
interference for imaginary couplings, the action still acquires a nonzero phase expectation, scaling with finite 
$L$ \cite{PRCim}.

\begin{figure}
  \begin{center}
    \includegraphics[width=6.5cm]{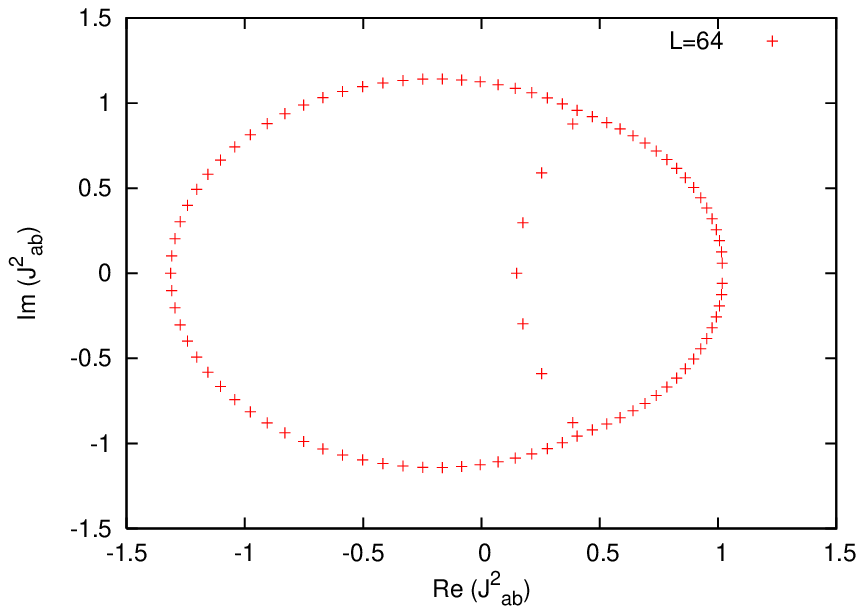}
    \includegraphics[width=6.5cm]{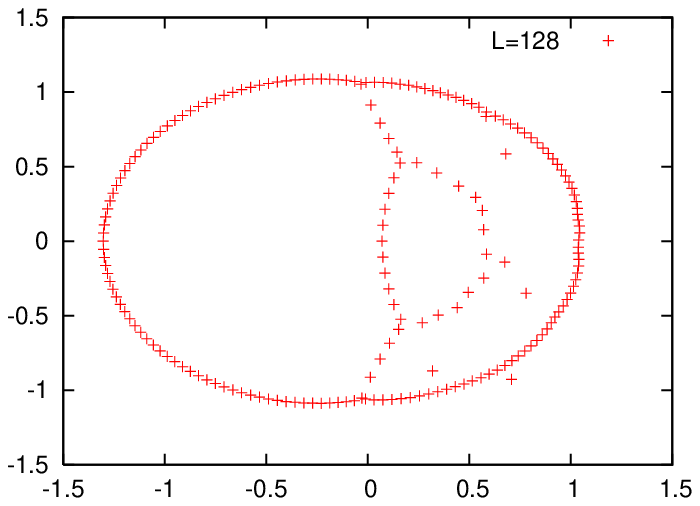}
    \includegraphics[width=6.5cm]{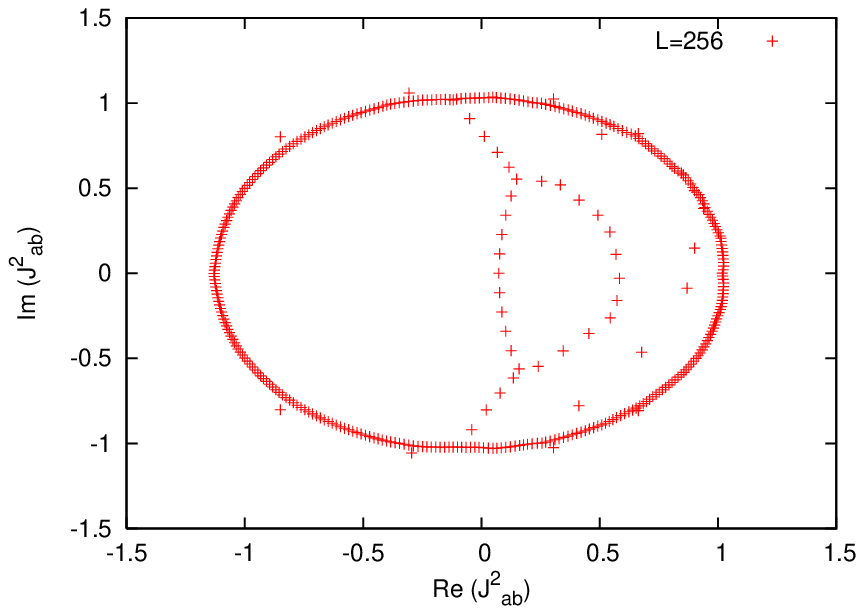}
    \includegraphics[width=6.5cm]{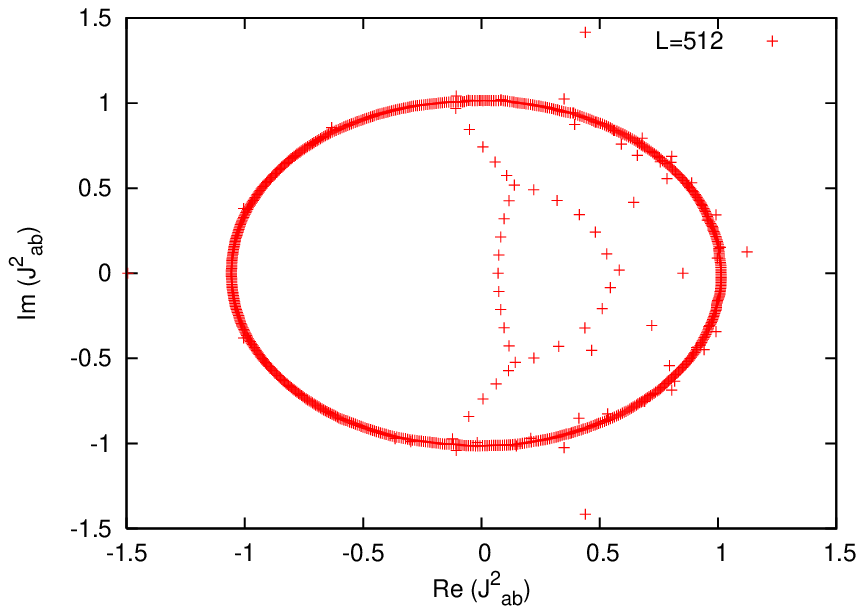}
  \end{center}
  \vspace{-13mm}
  \caption{Fisher zeroes in the complex $J_{ab}^{ \, 2}$ plane.}
\end{figure}

\section{LIGHT CONE \& INTEGRABILITY}
The correspondence of spin chains with light cone field theoretic treatments 
has been noted and studied via exact techniques such as the TBA 
\cite{light}\cite{str=fhm}\cite{mixrep}. 
For an explicit determination of twist-3 operators in QCD a further 
knowledge of the 
evolution at finite $N_c$ is required, however \cite{xxx=qcd}\cite{anomqcd}. Our Fisher zeroes scaling 
(under investigation) will now allow us to quantify this scaling in concert 
with the RG physics of the VBS states.

\end{document}